\documentclass[11pt]{article}

\usepackage[a4paper,margin=2.8cm]{geometry}
\usepackage{amsmath,amssymb,amsthm,mathtools}
\usepackage{hyperref}
\usepackage{enumitem}
\usepackage{placeins}
\usepackage{graphicx}
\newtheorem{theorem}{Theorem}[section]

\newtheorem{lemma}[theorem]{Lemma}

\theoremstyle{remark}
\newtheorem{remark}[theorem]{Remark}

\newcommand{\cH}{\mathcal H}
\newcommand{\cD}{\mathcal D}

\newcommand{\cL}{\mathcal L}
\newcommand{\lN}{\mathsf N}
\newcommand{\lD}{\mathsf D}
\newcommand{\Mas}{\operatorname{Mas}}
\newcommand{\Ker}{\operatorname{Ker}}

\newcommand{\sgn}{\operatorname{sgn}}
\newcommand{\dd}{\,\mathrm d}

\title{Zak Phase, Pole--Zero Invariants and the Maslov Index\\ of Bloch Cauchy Lines in One-Dimensional Periodic Media}
\author{Didier Felbacq\\[1mm]
\small Laboratoire Charles Coulomb, UMR 5221, Universit\'e de Montpellier and CNRS,\\
\small Place Eug\`ene Bataillon, 34095 Montpellier Cedex 05, France}
\date{}

\begin{document}

\maketitle

\begin{abstract}
The topological properties of the Bloch bands of a one-dimensional periodic medium are
described either by the Berry--Zak phase or by the poles and zeros of the surface
impedance. We show that both can be read from one geometric object: the line spanned
by the field and its flux (the Cauchy data) at an inversion centre. For each band, this
line turns as the quasi-momentum runs over the Brillouin zone. Its number of turns, the
Maslov index, is fixed by the parities (even or odd) of the Bloch functions at the two
band edges, and its parity is the Zak phase. In a gap, the Cauchy line of the solution
decaying on one side rotates monotonically with the energy, and an interface state between
two crystals is a crossing of two such lines. This gives the known criterion of at most
one state per gap in terms of the same lines. Examples with a stratified medium show that the
choice of inversion centre changes the interface spectrum. The last section discusses
what is missing to extend the construction to higher dimension.
\end{abstract}


\section{Cauchy data lines and the Maslov index}\label{sec:maslov}
When considering wave propagation in a one-dimensional periodic medium,
the topological properties of the bands can be
expressed as a holonomy, the Berry--Zak phase of the Bloch bundle
\cite{Zak1989,LinZhang2022}, or through the poles and zeros of the surface impedance
\cite{FelbacqRousseau2024}. In the following, we consider the propagation equation of a scalar Maxwell field
\begin{equation}
 -\frac{\dd}{\dd x}\!\left(\frac{1}{\mu(x)}\frac{\dd u}{\dd x}\right)
 = E\,\varepsilon(x)u,\qquad E=k_0^2,
 \label{eq:schrodinger-1d}
\end{equation}
where $\varepsilon,\mu$ are real, $1$-periodic functions satisfying
$0<c\le\varepsilon(x),\mu(x)\le C<\infty$ almost everywhere, possibly discontinuous at layer
interfaces, and $k_0$ is the vacuum wavenumber.

For $E_\parallel$
polarisation, $u=E_z$, the component of the electric field transverse to the structure, $\mu$ is the relative permeability and $\varepsilon$ the relative permittivity; the dual $H_\parallel$
polarisation is obtained by exchanging $\varepsilon$ and $\mu$. The flux is
$v=\mu^{-1}u_x$, where $u_x=\mathrm du/\mathrm dx$. Both $u$ and $v$ are continuous at an interface. The Cauchy vector $U=(u,v)^T$ obeys
\begin{equation}
 U_x=\begin{pmatrix}0&\mu(x)\\-E\varepsilon(x)&0\end{pmatrix}U,
 \qquad U(x)\in\mathbb R^2 \text{ for a real solution}.
 \label{eq:first-order}
\end{equation}
For real solutions we write $(q,p)=(u,v)$ for the coordinates of $U$, and equip $\mathbb R^2$ with
the symplectic form $\omega((q,p),(q',p'))=qp'-pq'$. Every real line $L$ is
Lagrangian, i.e. $\omega(U,U')=0$ for $U, U' \in L$ , and the set of lines is $\operatorname{Lag}(\mathbb R^2)=\mathbb{RP}^1$, the real projective line.
A line is written $\ell=\mathbb R(\cos\alpha,\sin\alpha)$, with angle $\alpha$
defined modulo $\pi$. The Neumann and Dirichlet lines are
$\lN=\mathbb R(1,0)$ and $\lD=\mathbb R(0,1)$, with angles $0$ and $\pi/2$.
The projective coordinate used in \cite{FelbacqRousseau2024} is
\begin{equation}
 \chi=\frac{q}{p}=\frac{u}{v}=\frac{u}{\mu^{-1}u_x},
  \label{eq:chi}
\end{equation}
therefore it holds
\begin{equation}
 \chi=0\iff\ell=\lD,\quad \chi=\infty\iff\ell=\lN.
\end{equation}
Zeros and poles of $\chi$ are the crossings of the Dirichlet and Neumann
lines. Up to a constant factor, $\chi$ is a surface impedance. Since $u$ and $v$
are continuous, the Wronskian identities below hold across discontinuities of
$\varepsilon$ and $\mu$.

\paragraph{Maslov index.}
Let $t\mapsto\ell(t)$, $t\in[t_0,t_1]$, be a continuous path of lines with a
continuous angle $\alpha(t)$. If $\ell_0$ is a fixed line, a crossing is a value
$t$ with $\ell(t)=\ell_0$; it is transverse if $\alpha$ is differentiable there
with $\alpha'(t)\ne0$, and its sign is $\sgn\alpha'(t)$. If all crossings are
transverse and none is at an endpoint, the Maslov index of the path relative to
$\ell_0$, written $\Mas(\ell_0,\ell)$, is the sum of the signs of the crossings.
For a loop $\ell$ ($\ell(t_1)=\ell(t_0)$) one has $\alpha(t_1)-\alpha(t_0)=m\pi$
with $m\in\mathbb Z$, and the number of signed crossings with any fixed line is
$m$; we put
\begin{equation}
  \Mas(\ell)=m=\frac{\alpha(t_1)-\alpha(t_0)}{\pi}
  =\deg\big(t\mapsto e^{2i\alpha(t)}\big).
  \label{eq:mas-loop}
\end{equation}
For two paths $\ell_1,\ell_2$ with angles $\alpha_1,\alpha_2$, the relative
Maslov index $\Mas(\ell_1,\ell_2)$ is the sum of the signs of
$(\alpha_2-\alpha_1)'$ over the crossings $\ell_1(t)=\ell_2(t)$, that is, the
values where $\alpha_2-\alpha_1\in\pi\mathbb Z$. With this orientation, a
positive crossing has increasing relative angle
\cite{BoossFurutani,LatushkinSukhtaiev}. Figure~\ref{fig:maslov} illustrates the definition on the Bloch lines of Section~\ref{sec:bands}.

\section{Periodic media: transfer matrix, band edges, Weyl lines}\label{sec:periodic}

Let $\varepsilon,\mu$ be even about $x=0$, and let $M(E)\in\mathrm{SL}(2,\mathbb R)$ be the transfer matrix over one period starting at the inversion centre $x=0$, i.e. it 
sends the Cauchy vector $U(0)$ to $U(1)$. Reflection acts on Cauchy data by
$\Pi=\operatorname{diag}(1,-1)$: if $u$ is a solution, so is $\tilde u(x)=u(-x)$,
and its Cauchy data are $(u(-x),-v(-x))$.

\begin{lemma}\label{lem:Mform}
\begin{equation}
  M(E)=\begin{pmatrix}a(E)&b(E)\\c(E)&a(E)\end{pmatrix},
  \qquad a(E)^2-b(E)c(E)=1.
  \label{eq:MonSym}
\end{equation}
Here $b(E)$ is the field $u_D(1;E)$ generated by the initial
Cauchy data $U_D(0)=(0,1)^T$, and $c(E)$ is the final flux of the
solution with $U_N(0)=(1,0)^T$.
\end{lemma}
\begin{proof}
The system \eqref{eq:first-order} has trace zero and $u,v$ are continuous
through the layers, so $\det M=1$. Let $R(x,y)$ be the propagator from $y$ to
$x$. The reflection property above gives $R(x,0)=\Pi R(-x,0)\Pi$, and
$R(-1,0)=R(0,-1)^{-1}$, with $R(0,-1)=R(1,0)=M$ by periodicity, so that $R(-1,0)=M^{-1}$. Hence $M=\Pi M^{-1}\Pi$, that is,
$\Pi M\Pi=M^{-1}$. For a matrix of determinant one, $M^{-1}$ has diagonal
entries $(d,a)$ while $\Pi M\Pi$ has diagonal entries $(a,d)$, so $a=d$.
\end{proof}

The bands are $\{|a|\le1\}$ and the open gaps are the connected components of
$\{|a|>1\}$. In a bounded open gap $b,c\neq0$, $bc=a^2-1>0$, and $a$ has a
constant sign. At an edge $E_*$ we have $|a|=1$, hence $bc=0$. We assume

\medskip
\noindent\textbf{(H)} $M(E_*)\neq\pm I$ at every band edge, i.e. all bounded gaps are open intervals.
\medskip

\noindent Then exactly one of $b,c$ vanishes at $E_*$, and $M(E_*)$ is a Jordan
block whose eigenline is $\lN$ if $c=0$ ($E_*$ is of \emph{Neumann type}, the
Bloch function is even) and $\lD$ if $b=0$ ($E_*$ is of \emph{Dirichlet type},
the Bloch function is odd). We write $s(E_*)=+1$ for Neumann type and $-1$ for
Dirichlet type, the inversion parity of the edge mode.

We therefore consider gaps that are genuine intervals. The non-generic case of a
closed gap ($M(E_*)=\pm I$, i.e.\ $b=c=0$, where two bands touch) is excluded.
At $E=0$, $M(0)=\left(\begin{smallmatrix}1&\int_0^1\mu\\0&1\end{smallmatrix}\right)$,
so the first edge is of Neumann type and satisfies (H).

For $E$ in a bounded open gap, let $\lambda_s(E)\in(-1,1)$ be the eigenvalue of
$M(E)$ of modulus less than one, and
\[
  L^s(E)=\Ker(M-\lambda_s),\qquad L^u(E)=\Ker(M-\lambda_s^{-1}).
\]

\begin{lemma}\label{lem:weyl}
Let $E$ range over a bounded open gap $G=(E_-,E_+)$.
\begin{enumerate}[label=(\alph*)]
\item $L^s(E)$ (resp.\ $L^u(E)$) is the Cauchy line at $0$ of the solutions
decaying as $x\to+\infty$ (resp.\ $x\to-\infty$), and $L^u=\Pi\,L^s$.
\item The angle $\alpha(E)$ of $L^s(E)$ is continuous on $[E_-,E_+]$ and
strictly increasing.
\item $L^s(E)\notin\{\lN,\lD\}$ for $E$ in the open gap. The two edges have
different types, and $\alpha$ increases by exactly $\pi/2$ across the gap.
\end{enumerate}
\end{lemma}

\begin{proof}
(a) If $U(0)\in\Ker(M-\lambda)$ then $U(n)=\lambda^nU(0)$, and
$M\Pi=\Pi M^{-1}$ gives $\Pi\Ker(M-\lambda)=\Ker(M-\lambda^{-1})$.

(b) An eigenvector is $U\propto(b,\lambda_s-a)$ (or $(\lambda_s-a,c)$ if $b=0$);
by (H) its direction depends continuously on $E$ up to the edges. For the
monotonicity let $\psi(x;E)$ be the solution decaying at $+\infty$, analytic in
$E$ in the open gap, $v=\mu^{-1}\psi_x$ and $\psi_E=\partial_E\psi$.
Differentiating \eqref{eq:schrodinger-1d} in $E$ gives
$-(\mu^{-1}\psi_{E,x})_x=E\varepsilon\psi_E+\varepsilon\psi$, and the Wronskian
$W=\psi\,\mu^{-1}\psi_{E,x}-v\,\psi_E$ satisfies $W'=-\varepsilon\psi^2$ on each
layer and is continuous at the interfaces. Since $\psi,\psi_E$ decay
exponentially, $W(+\infty)=0$ and
$W(0)=\int_0^\infty\varepsilon(x)\psi(x;E)^2\dd x>0$. Writing
$(\psi(0),v(0))=r(\cos\alpha,\sin\alpha)$ gives
\[\alpha'(E)=\frac{W(0)}{\psi(0)^2+v(0)^2}
=\frac{\int_0^\infty\varepsilon\psi^2\dd x}{\psi(0)^2+v(0)^2}>0.\]

(c) From the first row of $(M-\lambda_s)U=0$,
\[
  \chi_s=\frac{u}{v}=\frac{b}{\lambda_s-a}
  =-\,\sgn(a)\,\sgn(b)\sqrt{b/c},
\]
which is finite, non-zero and of constant sign in the open gap, since $b,c$ do
not vanish there and $b/c=bc/c^2>0$. Thus $\alpha$ stays in an open quadrant bounded by $\lN$ and
$\lD$. At the edges $\alpha$ is the angle of $\lN$ or $\lD$. Being strictly
increasing and confined to the quadrant, $\alpha$ goes from one bounding line
to the other, so the edge types differ and the rotation is $\pi/2$.
\end{proof}

\paragraph{Band orientation.}
For a band $B_n=[E_n^-,E_n^+]$, we shall also use two scalar periodic spectral facts:
\begin{equation}
 \sgn b(E)=(-1)^{n-1}\quad(E\in B_n),\qquad
 a(E_n^-)=(-1)^{n-1},\quad a(E_n^+)=(-1)^n.
 \label{eq:signb}
\end{equation}
Here $E_n^-$ and $E_n^+$ are the lower and upper energy edges of band $n$.
Consequently $E_n(\theta)$ increases on $[0,\pi]$ for odd $n$ and decreases
for even $n$. These are the usual alternation properties of Hill's equation
\cite{Eastham1973}. For the weighted equation \eqref{eq:schrodinger-1d} the
statements on $a$ and on the monotonicity of $E_n(\theta)$ are proved in
Appendix~\ref{app:alt}, which uses Lemma~\ref{lem:weyl}. To see the sign of $b$, start
from $b(0)=\int_0^1\mu>0$. Its zeros are simple Dirichlet eigenvalues:
at a zero $E_*$, the Wronskian identity gives
\[
 b'(E_*)v_D(1;E_*)=\int_0^1\varepsilon u_D^2\dd x>0.
\]
There is one such zero on each closed bounded gap (Lemma~\ref{lem:weyl}(c)) and none
inside a band, and $b$ changes sign at it since the zero is simple, so the sign changes
once between successive bands.

\section{Bloch bands: the loop of Cauchy lines}\label{sec:bands}

This section attaches to each band $n$ a closed loop of real lines in the Cauchy plane
and computes its Maslov index (Theorem~\ref{thm:bands}). A Bloch solution is complex, so
its Cauchy data $(u,v)$ at $0$ form a complex line, not a real one. We first show that,
after multiplying $v$ by $-i$, this complex line is the complexification of a real line;
this real line is the object whose turns are counted. The behaviour of the line at the
band edges, where one of $b,c$ vanishes, is part of the theorem.

\paragraph{The Bloch line.}
The bands are numbered $n=1,2,\dots$ from the bottom. The energy of band $n$ at
quasi-momentum $\theta\in\mathbb R/2\pi\mathbb Z$ is $E_n(\theta)$, with
$E_n(-\theta)=E_n(\theta)$, and $E=E_n(\theta)$ below. The Cauchy data at $0$ of a Bloch
solution $\psi(x+1)=e^{i\theta}\psi(x)$ lie in $\Ker(M(E)-e^{i\theta})$. Since
$\operatorname{tr}M=2a$, one has $a(E_n(\theta))=\cos\theta$ and
\begin{equation}
  b(E_n(\theta))\,c(E_n(\theta))=-\sin^2\theta .
  \label{eq:bc}
\end{equation}
For $\theta\notin\pi\mathbb Z$, $b$ and $c$ do not vanish, and the first and
second rows of $(M-e^{i\theta})U=0$ give the eigenvectors
\begin{equation}
  \begin{pmatrix}b\\ i\sin\theta\end{pmatrix},
  \qquad
  \begin{pmatrix}i\sin\theta\\ c\end{pmatrix}.
  \label{eq:bloch-vectors}
\end{equation}
Multiplying the second component by $-i$, that is, applying $A=\mathrm{diag}(1,-i)$,
gives the real vectors $(b,\sin\theta)$ and $(\sin\theta,-c)$, up to a nonzero complex
scalar. The transformed eigenspace is therefore the complexification of a unique real
line, which we call the \emph{Bloch line}:
\begin{equation}
  \ell_n(\theta)=\Big[A\,\Ker_{\mathbb C}\big(M(E_n(\theta))-e^{i\theta}I\big)\Big]\cap\mathbb R^2
  =\mathbb R\begin{pmatrix}b(E_n(\theta))\\ \sin\theta\end{pmatrix}
  =\mathbb R\begin{pmatrix}-\sin\theta\\ c(E_n(\theta))\end{pmatrix}.
  \label{eq:bloch-line}
\end{equation}
It is a real line, hence a Lagrangian line of $\mathbb R^2$. The factor $-i$ is the
natural choice because it removes the relative phase that the equation
$(M-e^{i\theta})U=0$ imposes between $u(0)$ and $v(0)$. For a solution with
$\psi(-x)=\overline{\psi(x)}$, that is, $S\psi=\psi$ with $S=RT$ the product of the
reflection $R\psi(x)=\psi(-x)$ and the complex conjugation $T\psi=\overline\psi$, the
Cauchy data have the form $(q,ip)$ with $q,p$ real.

\begin{theorem}[Zak--pole--zero--Maslov identity]\label{thm:bands}
Assume (H) at the edges of the bands $1,\dots,n$ (so that the gaps below band
$n$ are open). The line $\ell_n(\theta)$ of
\eqref{eq:bloch-line} extends to a continuous loop
$S^1\longrightarrow\mathbb{RP}^1$. Let $s_n^-,s_n^+\in\{\pm1\}$ be the
inversion parities of the Bloch modes at the \emph{lower} and \emph{upper}
energy edges of this band. With $\theta$ oriented increasingly and
$\omega((q,p),(q',p'))=qp'-pq'$, one has
\begin{equation}
 \Mas(\ell_n)=\frac{s_n^- -s_n^+}{2}\in\{-1,0,1\}.
 \label{eq:mas-formula}
\end{equation}
If $z_n$ is the parity of the number of Dirichlet-type edges of the band, then
\begin{equation}
 e^{i\gamma_{{\rm Zak},n}}=s_n^-s_n^+
   =(-1)^{\Mas(\ell_n)}=(-1)^{z_n}.
 \label{eq:zak-mas}
\end{equation}
\end{theorem}

\begin{remark}[Conventions]
The integer depends on the orientation of the coordinates $(u,v)$, on the
realification $A=\mathrm{diag}(1,-i)$ and on the orientation of $\theta$. Changing
any one of them reverses $\Mas(\ell_n)$ and leaves its parity unchanged.
\end{remark}

\begin{figure}[!ht]
\centering
\includegraphics[width=\textwidth]{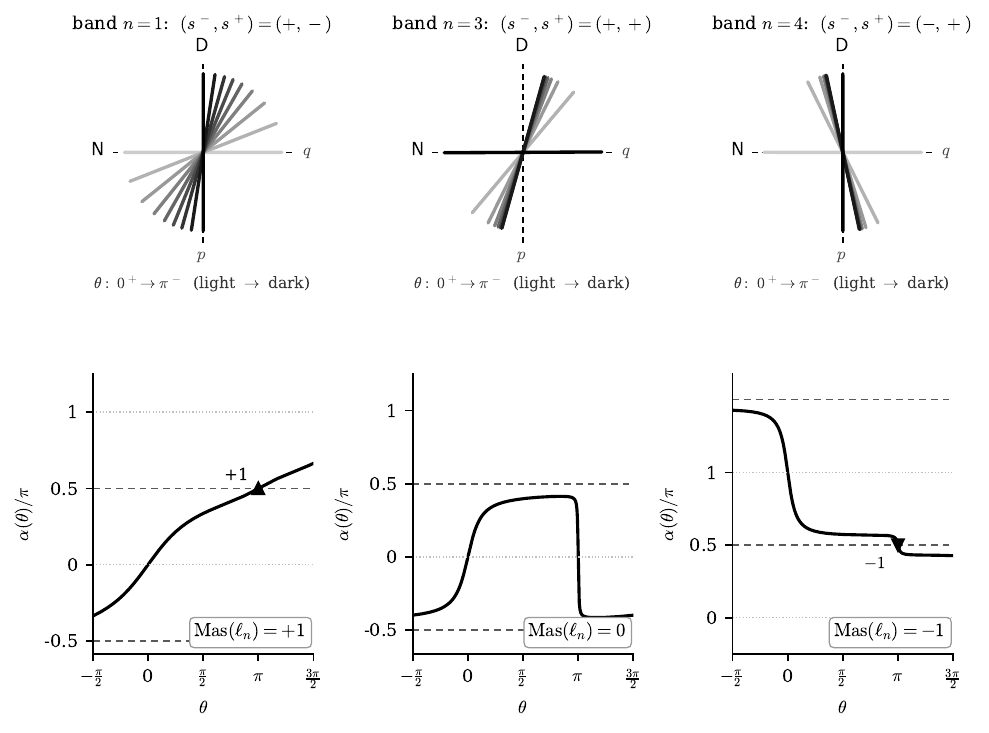}
\caption{Maslov index of the Bloch loop $\ell_n$ for three bands of the binary stack S1 of
Section~\ref{sec:numerics} ($\varepsilon_1=3.8$, $d_1=0.5$, inversion centre at the centre of
an $\varepsilon_2$ layer, denoted C2 in Section~\ref{sec:numerics}), $n=1,3,4$. Top: the real line $\ell_n(\theta)$ in the plane of Cauchy
data $(q,p)=(u,-iv)$ after realification, for $\theta\in(0,\pi)$, from light to dark; the dashed lines are
$\mathsf N=\mathbb R(1,0)$ and $\mathsf D=\mathbb R(0,1)$. Bottom: continuous angle $\alpha(\theta)$ of
the loop over one full period of $\theta$ (the part $\theta<0$ is deduced from $\theta>0$ by
$\ell_n(-\theta)=\mathrm{diag}(1,-1)\,\ell_n(\theta)$). Dashed horizontal lines: values of $\alpha$
for which the line is $\mathsf D$. Each crossing of $\mathsf D$ is marked ($\blacktriangle$:
$+1$, $\blacktriangledown$: $-1$, according to the direction in which $\alpha$ crosses). The Maslov index is the sum
of these signs, equivalently $[\alpha(\theta+2\pi)-\alpha(\theta)]/\pi$: it is $+1$, $0$ and $-1$ for
the edge parities $(s^-,s^+)=(+,-)$, $(+,+)$, $(-,+)$, in agreement with~\eqref{eq:mas-formula}.
The curve $\alpha(\theta)$ is plotted from the numerical $b(E_n(\theta))$.}
\label{fig:maslov}
\end{figure}

\begin{proof}
\emph{Extension to the edges.} In the open band $b,c\ne0$. At an edge
$\theta_*\in\{0,\pi\}$, $M(E_n(\theta_*))\ne\pm I$ and $bc=0$ by \eqref{eq:bc},
so exactly one of $b,c$ vanishes (they cannot both vanish, since $a=\pm1$ and
$b=c=0$ would give $M=\pm I$). Hence at each $\theta$ at least one of the two
vectors in \eqref{eq:bloch-line} is non-zero, and it spans $\ell_n(\theta)$.
If $c=0$ (Neumann type), the first expression
in \eqref{eq:bloch-line} tends to the non-zero vector $(b,0)$, which spans
$\lN$. If $b=0$ (Dirichlet type), the second tends to the non-zero vector
$(0,c)$, which spans $\lD$. The limit does not depend on the side of approach.
The limits at $\theta=-\pi$ and $\theta=\pi$ coincide since
$E_n(-\pi)=E_n(\pi)$. Hence $\ell_n$ is a continuous loop, and
$\ell_n(\theta_*)=\lN$ if $s=+1$ and $\lD$ if $s=-1$.

\emph{Crossings with $\lD$.} For $\theta\notin\pi\mathbb Z$ the first vector in
\eqref{eq:bloch-line} has first component $b(E_n(\theta))\ne0$, so $\ell_n(\theta)$
is not $\lD$ in the open band. Crossings of $\lD$ therefore occur only at
Dirichlet-type edges. There, with $\ell_n=\mathbb R(\cos\alpha,\sin\alpha)$,
\[
  \frac qp=\cot\alpha=-\frac{\sin\theta}{c(E_n(\theta))},
  \qquad
  \frac{\dd}{\dd\theta}\frac qp\Big|_{\theta_*}=-\frac{\cos\theta_*}{c(E_n(\theta_*))}\ne0,
\]
since $\sin\theta_*=0$. The crossing is transverse and, as $\dd(q/p)=-\dd\alpha$
at $\alpha=\pi/2$, its sign is $\sgn\alpha'(\theta_*)=\sgn[\cos\theta_*/c(E_n(\theta_*))]$.
By \eqref{eq:bc}, $b$ and $c$ have opposite signs in $(0,\pi)$ and do not
change sign there, so $\sgn c=-\sgn b$ near each edge, where $\sgn b=(-1)^{n-1}$ by
\eqref{eq:signb}. A Dirichlet-type edge at $\theta=0$ contributes
$-\sgn b$, and one at $\theta=\pi$ contributes $+\sgn b$. The index is the degree of the loop and does not depend on the base point; if the loop
crosses $\lD$ at $\theta=\pm\pi$, we parametrise it on $[-\pi+\delta,\pi+\delta]$ so that
the crossing is interior and counted once. Counting the signed crossings with $\lD$
gives, by \eqref{eq:mas-loop},
\[
  \Mas(\ell_n)=(-1)^{n-1}\Big(\mathbf 1_{s_\pi=-1}-\mathbf 1_{s_0=-1}\Big)
  =(-1)^{n-1}\,\frac{s_0-s_\pi}{2}.
\]
By \eqref{eq:signb}, the Bloch dispersion traverses each band from its lower to
its upper edge as $\theta$ runs from $0$ to $\pi$ for odd $n$, and in the reverse
order for even $n$. This is the usual alternation of periodic and antiperiodic
edge eigenvalues \cite{Eastham1973}, proved for the weighted problem in
Appendix~\ref{app:alt}.
For $n$ odd, $\theta=0$ is the lower edge and $s_0-s_\pi=s_n^--s_n^+$; for $n$
even, $\theta=0$ is the upper edge and $-(s_0-s_\pi)=s_n^--s_n^+$. This is
\eqref{eq:mas-formula}. It is nonzero exactly when the two edge types
differ, so $(-1)^{\Mas(\ell_n)}=s_n^-s_n^+=(-1)^{z_n}$.

\emph{Zak phase.} For the inversion-centred cell and the weighted scalar
product $\langle f,g\rangle_\varepsilon=\int_{-1/2}^{1/2}
\varepsilon\bar f g\dd x$, the inversion formula is
$e^{i\gamma_{{\rm Zak},n}}=s_0s_\pi$
\cite{Zak1989,LinZhang2022}. Since $\{s_0,s_\pi\}=\{s_n^-,s_n^+\}$,
this proves \eqref{eq:zak-mas}. A direct proof of this formula for the
present normalisation is given in Appendix~\ref{app:zak}.

\end{proof}

\begin{remark}[Pole--zero invariant]
In the interior of the band, the Cauchy coordinate of a Bloch solution is
$\chi_n=\psi(0)/v(0)=b/(i\sin\theta)$. At a Dirichlet-type edge
$\chi_n\to0$, and at a Neumann-type edge $\chi_n$ has a pole. Thus $z_n=1$ if
and only if the two edges are of type zero--pole or pole--zero
\cite{FelbacqRousseau2024}. Formula \eqref{eq:zak-mas} says that $z_n$ is the
parity of the number of times the Bloch line $\ell_n$ passes through $\lD$,
counted with sign. The integer $\Mas(\ell_n)$ is not a new classification: its parity
is the Zak phase. What the construction adds is that the band invariant and the
interface criterion of Section~\ref{sec:interface} are expressed with the same boundary
data. For periodic Jacobi operators, a formula for the Zak phase in terms of Weyl
functions is given in \cite{AmmariThalhammer2026}.
\end{remark}

\section{Gaps: interface states}\label{sec:interface}

Let $(\varepsilon_A,\mu_A)$ and $(\varepsilon_B,\mu_B)$ be two
real, positive, inversion-symmetric periodic structures satisfying (H), joined at
an inversion centre of each, which is taken as $x=0$: the medium $A$ occupies
$x<0$ and $B$ occupies $x>0$. The interface operator is
$\cL u=-\varepsilon^{-1}(\mu^{-1}u_x)_x$, self-adjoint on
$L^2(\mathbb R,\varepsilon\dd x)$ with $u$ and $v=\mu^{-1}u_x$ continuous at
the junction. Let $E$ be an eigenvalue of $\cL$ that lies in no band of
either medium and is not a band edge. Then $E$ lies in an open gap of each
medium, since a solution cannot be square integrable on a side whose medium has
$E$ in a band or at a band edge. We consider a bounded open gap $I$ common to the two media. For $E\in I$ put
\[
  L_+^s(E)=L^s_B(E),\qquad L_-^u(E)=L^u_A(E)=\Pi\,L_A^s(E).
\]
The standard matching argument gives
\begin{equation}
  E\in I\ \text{is an eigenvalue of }\cL
  \iff
  L_-^u(E)=L_+^s(E)
  \iff \omega(U_-^u,U_+^s)=0 ,
  \label{eq:interface-intersection}
\end{equation}
where $U_-^u,U_+^s$ are generators of the two lines. For a bounded open gap
$G=(E_-,E_+)$ of a medium let $t(G)=0$ if the lower edge $E_-$ is of Neumann
type and $t(G)=1$ if it is of Dirichlet type.

\begin{theorem}\label{thm:interface}
Let $G_A,G_B$ be bounded open gaps of $A$ and $B$ with
$I=G_A\cap G_B\ne\emptyset$.
\begin{enumerate}[label=(\roman*)]
\item $\cL$ has exactly one eigenvalue in $I$ if $t(G_A)+t(G_B)$ is odd, and none
if it is even.
\item Every crossing of the paths $E\mapsto L_-^u(E)$ and $E\mapsto L_+^s(E)$ is
positive. If the lower edges of $G_A,G_B$ are distinct, and likewise the upper
edges, there is no crossing at the end points of $I$, and the relative Maslov
index $\Mas(L_-^u,L_+^s)$ over $\overline I$ is the number of interface
states, $0$ or $1$.
\end{enumerate}
\end{theorem}

\begin{proof}
Let $\alpha_X$ be the angle of $L_X^s$. By Lemma~\ref{lem:weyl}, $\alpha_X$ is
strictly increasing, and $L_-^u=\Pi L_A^s$ has angle $-\alpha_A$. By
\eqref{eq:interface-intersection}, $E\in I$ is an eigenvalue if and only if
$\Phi(E):=\alpha_A(E)+\alpha_B(E)\in\pi\mathbb Z$. Since $\Phi$ is the
difference of the angles $\alpha_B$ and $-\alpha_A$ and is strictly increasing,
every crossing is positive, which is the first part of (ii).

On $G_X$ write $\alpha_X\equiv\tfrac\pi2t_X+x_X$ modulo $\pi$, with $t_X=t(G_X)$
and $x_X\in[0,\pi/2]$ strictly increasing from $0$ at the lower edge to $\pi/2$
at the upper edge (Lemma~\ref{lem:weyl}(c)). Then
$\Phi\equiv\tfrac\pi2(t_A+t_B)+x_A+x_B$ modulo $\pi$, and $x_A+x_B$ is strictly
increasing on $\overline I$.

Let $I=(E_a,E_b)$, where $E_a$ is the larger of the two lower edges and $E_b$
the smaller of the two upper edges. At $E_a$ one of $x_A,x_B$ is $0$. The other
is smaller than $\pi/2$, since $E_a<E_b$ is below the upper edge of its gap.
Hence $x_A+x_B<\pi/2$ at $E_a$, and in the same way $x_A+x_B>\pi/2$ at $E_b$.
Since $0\le x_A+x_B\le\pi$ on $\overline I$ and the sum is strictly increasing,
$0<x_A+x_B<\pi$ on $I$, and the value $\pi/2$ is taken exactly once. If
$t_A+t_B$ is even, $\Phi\equiv x_A+x_B\in(0,\pi)$ modulo $\pi$ on $I$, so there
is no solution. If it is odd, $\Phi\equiv\pi/2+x_A+x_B$, which vanishes modulo
$\pi$ if and only if $x_A+x_B=\pi/2$: exactly one solution. This proves (i).
For (ii), when the lower edges are distinct one of the two angles $x_X$ is $0$
and the other lies in $(0,\pi/2)$ at $E_a$, so $\Phi\not\in\pi\mathbb Z$ at
$E_a$, and likewise at $E_b$.
\end{proof}

\paragraph{From bands to gaps.}
Let $G_n$ be the gap above band $n$. Under (H), its lower-edge
type satisfies
\[
  t(G_n)\equiv \sum_{k=1}^{n}z_k+n-1\pmod 2.
\]
Indeed, the first band starts at a Neumann edge, passing through
band $k$ changes the edge type by $z_k$ modulo two, and crossing
each intervening gap changes it once more. Thus, by
Theorem~\ref{thm:interface}, two gaps $G_{n_A}$ and $G_{n_B}$
with nonempty intersection support exactly one interface state
if and only if
\begin{equation}
  \sum_{k=1}^{n_A}\Mas(\ell_k^A)
  +\sum_{k=1}^{n_B}\Mas(\ell_k^B)
  +n_A+n_B
  \equiv 1\pmod 2,
  \label{eq:bulk-boundary}
\end{equation}
and none otherwise. By Theorem~\ref{thm:bands},
$(-1)^{\Mas(\ell_k)}=e^{i\gamma_{{\rm Zak},k}}=(-1)^{z_k}$,
so this is also the criterion expressed through the Zak phases
or the pole--zero invariants of the bands below the two gaps.

Theorem~\ref{thm:interface}(i) and \eqref{eq:bulk-boundary} are the known
inversion-symmetric interface criterion of
\cite{XiaoZhangChan2014,ThiangZhang2023}, here written with the Maslov indices of
the band loops: the bulk quantity is the sum of the indices of the loops $\ell_k$ below
the gap, the boundary quantity is the number of crossings of the two Weyl lines in the
gap. The positivity of the crossings in part (ii) is the geometric form of the
monotonicity of Weyl functions (Lemma~\ref{lem:weyl}(b)).

\begin{remark}[Reflection phases]
For a homogeneous incident medium with Cauchy data $(1+r,i\beta_0(1-r))$, $\beta_0>0$,
the reflection coefficient of the semi-infinite stack \cite{BouchitteFelbacqZolla2005} is
$g=(i\beta_0\chi_s-1)/(i\beta_0\chi_s+1)$, $\chi_s=\cot\alpha$. This identifies the real
projective line with the unit circle, with $\lN\mapsto g=1$ and $\lD\mapsto g=-1$, and
$\Pi$ acts as complex conjugation. The condition $L_-^u=L_+^s$ is equivalent to
$g_Ag_B=1$, where $A$ is taken mirror-reversed. This is the matching of reflection
phases of \cite{XiaoZhangChan2014}, independent of $\beta_0$.
\end{remark}

\begin{remark}[Role of the symmetry]
The junction is at an inversion centre of each medium, which makes the edge lines
equal to $\lN$ or $\lD$. The two centres of a given crystal can give different edge
types for some bands, so $t(G)$ and the existence of a state depend on this choice
(Section~\ref{sec:numerics}).
\end{remark}

\section{Binary stacks: band loops and termination}\label{sec:numerics}
In this section we consider a structure whose period is made of two homogeneous slabs:
material 1, of thickness $d_1$, relative permittivity $\varepsilon_1$ and relative
permeability $\mu_1$, and material 2, of thickness $d_2=1-d_1$, with $\varepsilon_2$ and
$\mu_2$. The period is symmetric about the centre of each slab, so the structure has two
inversion centres per period, the centre of a slab of material 1 and the centre of a slab
of material 2. In a homogeneous slab $j$ the field is a combination of
$e^{\pm i\kappa_jx}$ with $\kappa_j=\sqrt{E\varepsilon_j\mu_j}$, and the transfer matrix over a
thickness $h$ is obtained from \eqref{eq:first-order} with constant coefficients. With
$Z_j=\mu_j/\kappa_j$,
\begin{equation}
 T_j(h;E)=\begin{pmatrix}
 \cos(\kappa_jh)&Z_j\sin(\kappa_jh)\\
 -Z_j^{-1}\sin(\kappa_jh)&\cos(\kappa_jh)
 \end{pmatrix},\qquad
 M=T_1(d_1/2)T_2(d_2)T_1(d_1/2).
 \label{eq:binary-monodromy}
\end{equation}
The cell starts at the centre of material 1 and is cut at the centre of the next slab of
material 1, so that $M$ has the form \eqref{eq:MonSym}, with $a$, $b$, $c$ its entries.
With $\phi_j=\kappa_jd_j$ the half-trace is
\[
 a=\cos\phi_1\cos\phi_2-\tfrac12(Z_1/Z_2+Z_2/Z_1)
 \sin\phi_1\sin\phi_2.
\]
The other inversion centre is obtained by interchanging materials 1 and 2.

We use S1: $\varepsilon_1=3.8$, $d_1=0.5$, and S2:
$\varepsilon_1=4.2$, $d_1=0.3$, with $\varepsilon_2=1$ and $\mu_1=\mu_2=1$, the structures
of \cite{FelbacqRousseau2026}. C1 denotes termination at the centre of material 1 and C2
at the centre of material 2. The band edges are the roots of $a^2=1$, their type is determined at each edge $E_*$ by $b(E_*)=0$ for Dirichlet type and $c(E_*)=0$ for Neumann type, and the degree of the loop is computed
from $\alpha=\arg(b+i\sin\theta)$. Interface states are counted from the sign changes
of $\omega(U_-^u,U_+^s)$ in each common gap. For all the bands below $k_0/2\pi=6$ and both
centres, the numerical degree equals $(s_n^--s_n^+)/2$. Table~\ref{tab:bands} and
Figure~\ref{fig:maslov} show the three possible values and their dependence on the
inversion centre.

\begin{table}[!ht]
\centering
\small
\begin{tabular}{llcccccc}
\hline
 & & $n=1$ & $n=2$ & $n=3$ & $n=4$ & $n=5$ & $n=6$\\
\hline
S1 & C1 & $(+,+)\!:0$ & $(-,-)\!:0$ & $(+,-)\!:1$ & $(+,+)\!:0$ & $(-,-)\!:0$ & $(+,-)\!:1$\\
S1 & C2 & $(+,-)\!:1$ & $(+,-)\!:1$ & $(+,+)\!:0$ & $(-,+)\!:-1$ & $(-,+)\!:-1$ & $(-,-)\!:0$\\
S2 & C1 & $(+,+)\!:0$ & $(-,+)\!:-1$ & $(-,-)\!:0$ & $(+,-)\!:1$ & $(+,+)\!:0$ & $(-,+)\!:-1$\\
S2 & C2 & $(+,-)\!:1$ & $(+,+)\!:0$ & $(-,+)\!:-1$ & $(-,-)\!:0$ & $(+,-)\!:1$ & $(+,+)\!:0$\\
\hline
\end{tabular}
\caption{Edge parities $(s_n^-,s_n^+)$ and $\Mas(\ell_n)=(s_n^--s_n^+)/2$ (rounded
numerical value, equal to the formula in all cases) for the first six bands of
the two structures and the two inversion centres.}
\label{tab:bands}
\end{table}

\paragraph{Interface.} We take S1 as medium $A$, S2 as medium $B$, and the four choices of
inversion centres. In each common bounded gap below $k_0/2\pi=6$ (Table~\ref{tab:iface}),
the number of eigenvalues found is $0$ or $1$ and equals the parity of $t(G_A)+t(G_B)$.

\begin{table}[!ht]
\centering
\small
\begin{tabular}{lcccc}
\hline
common gap, $k_0/2\pi$ ($n_A,n_B$) & C1--C1 & C1--C2 & C2--C1 & C2--C2\\
\hline
$[0.296,0.397]$ $(1,1)$ & 0 & 1 & 1 & 0\\
$[2.306,2.331]$ $(7,6)$ & 0 & 0 & 1 & 1\\
$[2.661,2.725]$ $(8,7)$ & 0 & 1 & 0 & 1\\
$[3.039,3.062]$ $(9,8)$ & 0 & 0 & 1 & 1\\
$[3.370,3.454]$ $(10,9)$ & 0 & 1 & 0 & 1\\
$[3.732,3.786]$ $(11,10)$ & 1 & 1 & 0 & 0\\
$[5.355,5.408]$ $(16,14)$ & 0 & 0 & 0 & 0\\
\hline
\end{tabular}
\caption{Number of interface states of S1$|$S2 in the common gaps, for the
inversion centres (S1--S2) of the two crystals.}
\label{tab:iface}
\end{table}

The table shows that the existence of the state depends on the choice of the
inversion centre of each crystal, as stated in the remark of
Section~\ref{sec:interface}. In particular, for the junction at the centres of
the $\varepsilon_2$ layers (C2--C2) there is a state in the gaps near
$k_0/2\pi\simeq2.3$ and $2.7$, whereas there is none in these two gaps for the
junction at the centres of the $\varepsilon_1$ layers (C1--C1).

\FloatBarrier
\section{Higher dimension: what requires an additional construction}\label{sec:higher}

\paragraph{Known framework.} Let $\Omega\subset\mathbb R^d$ be a bounded domain with
smooth boundary $\Gamma$, let $\cH_\Gamma=H^{1/2}(\Gamma)\oplus H^{-1/2}(\Gamma)$ be the real trace space with
the symplectic form $\omega((f,g),(f',g'))=\langle g',f\rangle_\Gamma-\langle g,f'\rangle_\Gamma$,
and let $\cD=\{0\}\oplus H^{-1/2}(\Gamma)$. The space $\Lambda(\lambda)$ of Cauchy data of
the solutions of $(-\Delta-\lambda)u=0$ in $H^1(\Omega)$ is a Lagrangian subspace forming a
Fredholm pair with $\cD$, and it is the graph of the Dirichlet-to-Neumann operator away
from Dirichlet eigenvalues. The Maslov index of $\lambda\mapsto\Lambda(\lambda)$ relative to
$\cD$ counts eigenvalues \cite{LatushkinSukhtaiev,Furutani2004}. This is the analogue of
Lemma~\ref{lem:weyl}(b); it does not involve Bloch bands.

\paragraph{Three differences.}
\begin{enumerate}[label=(\arabic*)]
\item \emph{The band object.} In dimension one, the Bloch line is a line of $\mathbb R^2$,
hence Lagrangian, and $\operatorname{Lag}(\mathbb R^2)=\mathbb{RP}^1$ is a circle: the
angle $\alpha$ gives the degree of a loop. For $d\ge2$, the traces of the eigenspace of a simple band
form a complex line in an infinite-dimensional space. This line is not Lagrangian,
there is no angle coordinate, and a Maslov index is defined only relative to a
reference subspace forming a Fredholm pair with the path. A Lagrangian object attached to one band requires a choice of
boundary cut, of real structure (the analogue of $A=\operatorname{diag}(1,-i)$) and of
reference subspace.
\item \emph{The Brillouin torus.} In dimension one the Brillouin zone is a circle with
two inversion-invariant points, and every isolated band gives a loop. For $d\ge2$ the
zone is a torus with $2^d$ inversion-invariant momenta, and bands can overlap or be
degenerate. One considers inversion-invariant closed paths $C$ through two
inversion-fixed momenta, along which the band is simple and isolated, in the symmetry
class where the Berry phase is quantised. The comparison depends on $C$, on the cell
convention and on the realification.
\item \emph{Interfaces.} For a planar interface the tangential momentum $k_\parallel$
is a parameter. At fixed $k_\parallel$, the spaces of Cauchy data at the interface of solutions
decaying into each medium
replace $L_-^u$ and $L_+^s$, and an interface mode is an intersection whose dimension is
its multiplicity. The zero-or-one count of Theorem~\ref{thm:interface} uses a rotation of
exactly $\pi/2$ across the gap and edges of opposite types, which rest on
$\operatorname{Lag}(\mathbb R^2)$ being a circle; it gives no such bound in higher
dimension. At fixed energy in a common gap, a path in $k_\parallel$ gives a spectral
flow, which differs from the number of modes at fixed momentum. The
Maslov--spectral-flow framework \cite{LatushkinSukhtaiev} applies once the Lagrangian and
Fredholm properties are established for these half-space problems.
\end{enumerate}

The one-dimensional construction thus brings together two features that must be
distinguished in higher dimension: selecting a single Bloch band and obtaining a
Lagrangian space of boundary data. The trace of a simple Bloch eigenspace retains
the band information, whereas the full Cauchy-data space supplies the Lagrangian
structure. Extending the band-loop picture would require a construction that
connects these two objects. Only after such a construction is specified can its
index be compared with the Berry phase along a symmetry-compatible loop. On the
interface side, the decaying Cauchy-data spaces provide a more direct starting
point for an intersection description, although the scalar quarter-turn argument
no longer supplies a zero-or-one count. These distinctions identify the scope of
the present geometric picture and the additional work needed to extend it.

\section{Conclusion}
Cauchy data give a common description of band topology and interface states in
one-dimensional inversion-symmetric media. The Zak and pole--zero invariants are
the parity of the Maslov index of the realified Bloch line. In a common gap,
interface states occur where the two decaying Cauchy lines meet; their monotonic
rotation gives the zero-or-one criterion. The examples show how changing the
termination can create or remove such a state. In higher dimensions, the main
unresolved step is to associate a Lagrangian space of boundary data with an
individual Bloch band.

\appendix
\section{Sign of \texorpdfstring{$a$}{a} and dispersion in a band}\label{app:alt}
\begin{lemma}[Sign of $a$ and dispersion in a band]
Assume (H) at the edges of the bands $1,\dots,n$.
\begin{enumerate}[label=(\alph*)]
\item In the gap $G_k$ above band $k\le n-1$, $\sgn a=(-1)^k$.
\item In the open band $n$, $a(E)=\cos\theta(E)$ with $\theta(E)\in(0,\pi)$ and
$\theta'(E)\ne0$. Hence $a$ is strictly monotone in the band, $a=(-1)^{n-1}$ at its
lower edge and $a=(-1)^n$ at its upper edge, and $E_n(\theta)$ is strictly
monotone on $[0,\pi]$: increasing if $n$ is odd, decreasing if $n$ is even.
\end{enumerate}
\end{lemma}
\begin{proof}
(a) In $G_k$, $\sgn\chi_s=-\sgn a\,\sgn b$ by the formula in the proof of
Lemma~\ref{lem:weyl}(c). By Lemma~\ref{lem:weyl}(c), $\alpha$ goes from the lower
edge line to the upper edge line, so $\chi_s=\cot\alpha>0$ if $t(G_k)=0$
(from $\lN$ to $\lD$) and $\chi_s<0$ if $t(G_k)=1$ (from $\lD$ to $\lN$). Thus
$\sgn\chi_s=(-1)^{t(G_k)}$. The function $b$ vanishes in $\overline{G_k}$ only at
the Dirichlet-type endpoint, where it changes sign (see \eqref{eq:signb}), and its
sign in band $k$ is $(-1)^{k-1}$. Hence $\sgn b|_{G_k}=(-1)^{k-1+t(G_k)}$: if the lower
edge is of Dirichlet type, $b$ changes sign at it, and otherwise it keeps the sign of
band $k$ up to the upper edge. Therefore
$\sgn a=-(-1)^{t(G_k)}(-1)^{k-1+t(G_k)}=(-1)^k$.

(b) For $E$ in the open band, $|a|<1$ and $\theta(E)=\arccos a(E)\in(0,\pi)$ is
differentiable. Let $\psi(x;E)$ be a Bloch solution with multiplier $e^{i\theta(E)}$,
with Cauchy data at $0$ differentiable in $E$, and $v=\mu^{-1}\psi_x$. Such a choice exists: $M(E)$ is analytic in $E$ and $e^{i\theta(E)}$ is a simple eigenvalue, since $e^{-i\theta}\ne e^{i\theta}$ for $\theta\notin\pi\mathbb Z$, so the eigenvector can be chosen analytic in $E$. The quantity
$F=\operatorname{Im}(\bar\psi v)$ is independent of $x$, and $F\ne0$. Indeed, if $F=0$, then $\bar\psi(0)v(0)$ is real, so $U(0)=e^{i\varphi}U_r$ with $U_r$ real; then $U(1)=M(E)U(0)=e^{i\varphi}M(E)U_r$ with $M(E)U_r$ real, and the Bloch condition $U(1)=e^{i\theta}U(0)$ forces $e^{i\theta}\in\mathbb R$, that is $e^{i\theta}=\pm1$, which is excluded. We put $V=\bar\psi\,\partial_Ev-\bar v\,\partial_E\psi$. The
equations give $V'=-\varepsilon|\psi|^2$. Differentiating
$\psi(1;E)=e^{i\theta(E)}\psi(0;E)$ in $E$ gives
$V(1)-V(0)=-2\theta'(E)F$, so
\[
 \theta'(E)=\frac{1}{2F}\int_0^1\varepsilon(x)|\psi(x;E)|^2\dd x\ne0 .
\]
Thus $a$ is strictly monotone in the band. At the lower edge of band $n$, $a=1$ if
$n=1$ (the constant function is a periodic solution at $E=0$) and $a=(-1)^{n-1}$
if $n\ge2$ by (a) applied to $G_{n-1}$. Since $|a|=1$ at both edges and $a$ is
strictly monotone, $a=(-1)^n$ at the upper edge. The statement on $E_n(\theta)$
follows from $a(E_n(\theta))=\cos\theta$.
\end{proof}

\section{The Zak phase for the present normalisation}\label{app:zak}
The operators $R$, $T$ and $S=RT$ commute with the equation
\eqref{eq:schrodinger-1d}, because $\varepsilon,\mu$ are real and even, so they
map Bloch solutions to Bloch solutions ($T$ changes $\theta$ into $-\theta$).
We use the unit cell $[-\tfrac12,\tfrac12]$, which $R$ preserves,
and the energy-weighted scalar product
$\langle f,g\rangle_\varepsilon=\int_{-1/2}^{1/2}
\varepsilon(x)\overline{f(x)}g(x)\dd x$, antilinear in the first
argument. The Zak phase is
$\gamma=i\int_{-\pi}^{\pi}\langle u_\theta,\partial_\theta u_\theta\rangle_\varepsilon\dd\theta$
for a continuous periodic gauge, $\psi_{-\pi}=\psi_{\pi}$, where
$u_\theta=e^{-i\theta x}\psi_\theta$ is the periodic part. In terms of the periodic parts the
gauge condition reads $u_{-\pi}(x)=e^{2\pi ix}u_{\pi}(x)$, since $e^{i\pi x}\psi_{-\pi}=e^{i\pi x}\psi_\pi=e^{2\pi ix}\,e^{-i\pi x}\psi_\pi$. The sign convention for
$\gamma$ plays no role below, since $e^{i\gamma}=\pm1$ (reversing the orientation of
$\theta$ changes $\gamma$ into $-\gamma$).
\begin{enumerate}[label=\textit{(\alph*)},nosep]
\item \emph{An $S$-fixed section on $[0,\pi]$.} We choose a continuous real generator
$(q_\theta,p_\theta)$ of $\ell_n(\theta)$ on $[0,\pi]$, which is possible on an
interval. Let $\psi_\theta$ be the Bloch solution with Cauchy data
$A^{-1}(q_\theta,p_\theta)=(q_\theta,ip_\theta)$ at $0$, normalized in $L^2_\varepsilon$ of
the cell and with continuous dependence on $\theta$. Then $S\psi_\theta$ has
Cauchy data $(\overline{q_\theta},-\overline{ip_\theta})=(q_\theta,ip_\theta)$, so
$S\psi_\theta=\psi_\theta$.
\item \emph{Extension to $[-\pi,0]$.} We put $\psi_\theta=T\psi_{-\theta}$ for
$-\pi\le\theta<0$. It is a normalized Bloch solution with multiplier $e^{i\theta}$,
and it is $S$-fixed because $ST=TS$.
\item \emph{Jumps at the fixed points.} At $\theta_*\in\{0,\pi\}$, $\psi_{\theta_*}$
is an edge mode, so $R\psi_{\theta_*}=s_{\theta_*}\psi_{\theta_*}$ with
$s_{\theta_*}=\pm1$. Then $S\psi_{\theta_*}=\psi_{\theta_*}$ gives
$T\psi_{\theta_*}=R\psi_{\theta_*}=s_{\theta_*}\psi_{\theta_*}$. Hence
$\psi_{0^-}=s_0\,\psi_{0^+}$ and $\psi_{-\pi}=s_\pi\,\psi_{\pi}$. We define
$\psi'_\theta=\psi_\theta$ for $\theta>0$ and $\psi'_\theta=s_0\psi_\theta$ for
$\theta<0$. This section is continuous on $[-\pi,\pi]$ and
$\psi'_{-\pi}=s_0s_\pi\,\psi'_{\pi}$.
\item \emph{The Berry connection vanishes.} For the periodic parts,
$S$ acts as $u\mapsto\overline{u(-x)}$, independently of $\theta$, and fixes
$u'_\theta$ and $\partial_\theta u'_\theta$. It is antiunitary for the
$\varepsilon$-weighted inner product ($\varepsilon$ is real and even), so
$\langle u',\partial_\theta u'\rangle_\varepsilon$ equals its complex conjugate
and is real. Normalization gives
$\mathrm{Re}\langle u',\partial_\theta u'\rangle_\varepsilon=0$.
Hence $\langle u'_\theta,\partial_\theta u'_\theta\rangle_\varepsilon=0$ on
$(-\pi,0)\cup(0,\pi)$.
\item \emph{Closing the gauge.} We write $s_0s_\pi=e^{i\chi}$ with $\chi\in\{0,\pi\}$,
and put $\psi''_\theta=e^{i\chi(\theta+\pi)/2\pi}\psi'_\theta$. Then
$\psi''_{\pi}=e^{i\chi}\psi'_{\pi}=\psi'_{-\pi}=\psi''_{-\pi}$, so $\psi''$ is a
continuous periodic gauge, still normalised since the factor has modulus one, and
$\langle u''_\theta,\partial_\theta u''_\theta\rangle_\varepsilon=i\chi/2\pi$. Therefore
$\gamma=i\int_{-\pi}^{\pi}(i\chi/2\pi)\dd\theta=-\chi$, and
$e^{i\gamma_{{\rm Zak},n}}=e^{-i\chi}=s_0s_\pi=s_n^-s_n^+$, the last equality
because $\{s_0,s_\pi\}=\{s_n^-,s_n^+\}$.
\end{enumerate}

%

\end{document}